\documentclass[12pt]{iopart}
\newcommand{\beq}{\begin{equation}}
\newcommand{\eeq}{\end{equation}}

\def\ep{\epsilon}

\def\jj{{\mbox{${\cal J}$}} {}}

\def\beq{\begin{equation}}
\def\ee{\end{equation}}
\def\lsim{\mathrel{\rlap{\lower4pt\hbox{\hskip1pt$\sim$}}
    \raise1pt\hbox{$<$}}}
\def\gsim{\mathrel{\rlap{\lower4pt\hbox{\hskip1pt$\sim$}}
    \raise1pt\hbox{$>$}}}

\def\jb{{\mbox{\boldmath ${\cal J}$}} {}}

\def\bfQ{{\bf Q}}

\def\bfB{{\bf B}}
\def\bfs{{\bf s}}

\def\ts{\times}

\def\curl{\nabla {\ts}}

\def\bfv{{\bf v}}

\def\bfV{{\bf V}}

\def\bfB{{\bf B}}

\def\div{\nabla\cdot}

\begin{document}


\title{On deriving flux freezing in magnetohydrodynamics  by direct differentiation}

\author{Eric G. Blackman}

\address{Department of Physics and Astronomy, University of Rochester, Rochester, NY 14627}
\ead{blackman@pas.rochester.edu}
\begin{abstract}
The magnetic flux freezing theorem
 is a  basic principle of ideal magnetohydrodynamics (MHD), a  commonly used approximation to describe the  aspects of  astrophysical and  laboratory plasmas. The theorem states that the magnetic flux---the integral of magnetic field penetrating a surface---is conserved in time  as that surface is distorted in in time by   fluid motions.   Pedagogues of MHD commonly derive flux freezing  without  showing  how to take the material derivative of a general flux integral and/or assuming a vanishing field divergence from the outset.    Here I  avoid these shortcomings and derive flux freezing by direct differentiation, explicitly using a Jacobian to transform between the evolving field-penetrating surface  at different times.  The approach is  instructive for its generality and  helps elucidate the role of magnetic monopoles in  breaking flux freezing.  The paucity of appearances of this derivation  in standard MHD  texts suggests  that  its  pedagogic value is underappreciated.
\end{abstract}

\bigskip
accepted by {European  Journal  of  Physics}
\maketitle

\section{Introduction}

Magnetohydrodynamics (MHD) is the simplest generalization of hydrodynamics for a  sufficiently ionized collisional plasma \cite{alfven43,chandra61,parker79}.  The relative motion between positive and negative  charge carriers creates currents which can sustain magnetic fields and electric fields are generated by induction. Charge separation and plasma oscillations are assumed to occur on  small enough spatial and temporal scales that the plasma is considered to be neutral on  macroscopic scales of interest. 
As in hydrodynamics, a high rate of particle interactions ensures that deviations from Maxwellian velocity distributions are small.  

Non-relativistic MHD  limit has been a mainstay of theoretical astrophysics 
since most of the material inside stars and between them   is composed of non-relativistic magnetized plasma and is commonly treated in the MHD approximation. The limit is also widely used in approximating the  dynamics and stability of fusion device plasmas.
Students in  astrophysics and plasma physics are typically  exposed to 
MHD either in advanced undergraduate  or graduate courses.
 
The solution of physical problems in MHD  requires equations for  mass conservation equation, momentum conservation, energy  evolution equation  and the magnetic induction equation.
 The latter is the subject of the present paper 
  and  is given in CGS units by \cite{parker79}
\beq
{\partial {\bf B}\over \partial t}  =\curl ({{\bf V} \times {\bf B}}) - \curl (\nu_M \curl {\bf B}), 
\label{1}
\eeq
where $\bf B$ is the magnetic field, $\bf V$ is the plasma velocity, and $\nu_M\equiv{ \eta c^2\over 4\pi}$ is the magnetic diffusivity in terms of the resistivity $\eta$ and speed of light $c$.   The form of Eqn (1) is identical to that of vorticity evolution in incompressible gravitational hydrodynamics 
if $\bf B$ is replaced by  vorticity $\mbox{\boldmath$\omega$} \equiv \curl {\bf V}$  and $\nu_M$ is replaced by the kinematic viscosity $\nu$.

Eq (\ref{1}) is  derived by starting with   Faraday's law ${\partial {\bf B} \over \partial t}=c \curl {\bf E}$, where $\bf E$ is the electric field. 
Eliminating $\bf E$ in terms of $\bf B$ and $\bf J$ is  then accomplished by use of
   Ohm's law, ${\bf E} + {\bf V} \times {\bf B} = \eta{\bf J}$, where $\bf J$ is the current density. Use of the non-relativistic Amp\`eres law $\curl {\bf B} = {4\pi \over c}{\bf J}$ then relates   $\bf J$ and $\bf B$. The aforementioned Ohm's law is  derived by subtracting the separate momentum density equations for positive and negative charge carriers. The resistive term  arises in Ohm's law when there are finite but small deviations from 
 Maxwellian distributions of the charge carriers.
   That these deviations are assumed to be weak highlights that the relevance of MHD when
    many collisions between charged particles occurs over 
   dynamical times of interest.  Collisionless  plasmas  have more complicated Ohm's laws.
   
  The induction equation of  {\it ideal MHD} corresponds to Eq. (\ref{1}) when the $\nu_M$ term is neglected.   The ratio of magnitudes of the  second term to the third term in Eq. $(\ref{1})$ can be approximated by  the magnetic Reynolds number $R_M \equiv {VL\over \nu_M}$,  where $V$ is the velocity magnitude and $L$ is the characteristic gradient scale of velocity or magnetic field. When $R_M>>1$, the resistive term is often ignored and ideal MHD assumed.  Actually, many astrophysical plasmas are turbulent which implies a  spectrum  of 
eddies of different scales and energies.  Such flows always have a  microphysical dissipation scale at which  $R_M=1$ so in practice one must think about $R_M$ as a scale dependent quantity.
Subtleties involved with applying the limit of  ideal MHD  to turbulent flows are  discussed in \cite{bf08}
and \cite{eyink06}.   For the laminar  (non-turbulent)
case, the limit $R_M>>1$ leads more  straightforwardly to the ideal MHD approximation and  that is the focus of this paper.

The ideal MHD limit   is indeed the limit for which   "magnetic flux freezing"  or  Alfv\'en's theorem holds.  This theorem (analogous to the Kelvin circulation  theorem of ideal hydrodynamics \cite{kelvin1869,landau87}  in which vorticity flux is frozen) states that the magnetic flux through a material surface is conserved even as velocity  flows distort that surface in time.   Mathematically this means 
\beq
{D\Phi_B \over Dt } =0,
\label{2}
\ee
where
 $D/Dt$ indicates the material  or Lagrangian derivative ${\partial\over \partial  t }+
  {\bf V}({\bf x},t)\cdot \nabla$ for space and time dependent flow velocity ${\bf V}({\bf x},t)$ and magnetic  flux  $\Phi_B \equiv \int {\bf B} \cdot d{\bf S} $, where the integral represents a surface integral
 over an open surface. The rest of this paper addresses   derivations of Eq. (\ref{2}).

 In section 2, I  derive Eq. (\ref{2}) by a  formal  material derivative of the flux
 integral. In section 3, I  compare this derivation to  other derivations 
  commonly found in standard  texts.  In section 4,  I  address why  the derivation of section 2 facilitates a better physical understanding of the role magnetic monopoles play in  violating flux freezing compared to the other derivations discussed in section 3 . I conclude in section 4.

\section{Material derivative of the flux and  derivation of flux freezing}

Consider  an open surface within a plasma at $t=0$ through which  magnetic field lines penetrate. The surface has  a differential area $dS_0=d\sigma_1d\sigma_2$, where $\sigma_1$ and $\sigma_2$ define local Cartesian coordinates.    Via the distorting action of a smoothly varying time and space dependent velocity flow $\bf V$, this surface  evolves to a new surface  with differential $d\bf S$  at   time $t\ge0$.   To compute $D\Phi_B/ Dt$  
we must then account for the fact that both $\bf B$  and $d{\bf S}$  can depend on  space and time.

The time-evolved surface measure $d\bf S$ can be related to the initial  surface measure by a Jacobian  transformation. With this in mind, let us define a general flux $\Phi_Q\equiv \int \bfQ\cdot d{\bf S}$
for an arbitrary vector $\bfQ$, not necessarily divergence free.   Then
\beq
{D\Phi_Q\over Dt}={D\over Dt}\int\bfQ\cdot d{\bf S}=
{D\over Dt}\int\bfQ\cdot \jb d{ S}_0,
\label{j1}
\eeq
where $d{\bf S}$ is   the differential surface area vector at  an arbitrary $t\ge 0$ and $dS_0=||d{\bf S}_0||$  is the scalar differential surface area of the surface at  $t=0$. Here $\jb$ is the Jacobian vector relating $d{\bf S}$ to a   coordinate transformation of $d{\bf S}_0$, and has components
\beq
\jj_q=\ep_{qrs}{\partial x_r\over \partial \sigma_1}{\partial x_s\over \partial \sigma_2},
\label{a1}
\eeq
where $\ep_{qrs}$ is the Levi-Civita symbol and 
$x_1, x_2,x_3$ are the local Cartesian coordinates of the evolving surface element $d\bf S$.
 Three coordinates are required for the evolving surface  as the surface normal can evolve away from its initial direction.

 Because the right side of (\ref{j1}) now  involves an integral over a fixed surface, we can take the $D\over Dt$ inside the integral to obtain
\beq
{D\over Dt}\int\bfQ\cdot \jb d{S}_0=
\int \left(\jb\cdot {D\over Dt}\bfQ + \bfQ\cdot {D\jb\over Dt}\right)  d{S}_0.
\label{j3}
\eeq
We now need an expression for $D\jb/Dt$. 
Changing the indices $q,r,s$ in (\ref{a1}) to $k,i,j$ and taking the material derivative gives
\beq
{D\jj_k\over Dt}=\ep_{kij}\left[{\partial( Dx_i/Dt )\over \partial \sigma_1}{\partial x_j \over \partial \sigma_2}
+{\partial x_i \over \partial \sigma_1}{\partial (Dx_j/Dt )\over \partial \sigma_2}\right].
\label{a2}
\eeq
Using
${Dx_j\over Dt}= V_j$;  ${Dx_i\over Dt}= V_i$;  ${\partial V_i\over \partial \sigma_1}={\partial V_i \over \partial x_m}{\partial x_m \over\partial \sigma_1}
$;
and 
${\partial V_j\over \partial \sigma_2}={\partial V_j \over \partial x_m}{\partial x_m\over  \partial \sigma_2}
$
 in Eq. (\ref{a2}) then gives
\beq
{D\jj_k\over Dt}=\ep_{kij}\left[{\partial V_i \over \partial x_m}{\partial x_m \over \partial \sigma_1}{\partial x_j \over \partial \sigma_2}
+{\partial x_i \over \partial \sigma_1}{\partial V_j \over \partial x_m}{\partial x_m \over \partial \sigma_2}\right].
\label{a3}
\eeq
Since $\ep_{kij}=-\ep_{kji}$,  we can  interchange  indices  $i$ and $j$ in (\ref{a3}) to  obtain
\beq
\begin{array}{r}
{D\jj_k\over Dt}=\ep_{kij}{\partial V_i \over \partial x_m}\left[
{\partial x_m \over \partial \sigma_1}{\partial x_j \over \partial \sigma_2}
-{\partial x_j\over \partial \sigma_1}{\partial x_m \over \partial \sigma_2}\right]
= \ep_{kij} {\partial V_i \over \partial x_m} \ep_{mjq}\jj_q\\
=\jj_k \nabla\cdot \bfV - \jj_i{\partial V_i\over \partial x_k },
\label{a4}
\end{array}
\eeq
where the second equality follows from using (\ref{a1}) multiplied by $\epsilon_{kmn}$
and the third equality follows from using $\epsilon_{kij}\epsilon_{mjq}=
\delta_{qk}\delta_{mi}-\delta_{qi}\delta_{mk}$.

Using (\ref{a4}), along with the definition ${D/Dt}$ and the fact that $(\bfV\cdot\nabla \bfQ)_i=\bfV\cdot\nabla Q_i$ in Cartesian coordinates, we now obtain for $(\ref{j3})$
\beq
{D\over Dt}\int\bfQ\cdot \jb d{S}_0 =
\int\left({\partial Q_i\over \partial t}+\bfV\cdot\nabla Q_i  +   Q_i\nabla\cdot\bfV-\bfQ\cdot \nabla V_i\right)\jj_i d{S}_0. 
\label{j5}
\eeq
Using $\jj_i d{S}_0= dS_i$ and the vector identity 
\beq
\curl(\curl \bfQ)= \bfQ\cdot \nabla\bfV-\bfV\cdot \nabla\bfQ-\bfQ\div\bfV+\bfV \div\bfQ, 
\label{j6}
\eeq
Eq. (\ref{j5}) becomes
\beq
{D\over Dt}\int \bfQ\cdot d{\bf S} =
\int\left({\partial {\bf Q}\over \partial t}-\curl(\bfV\times \bfQ)  +   \bfV \div\bfQ \right)\cdot d{\bf S}. 
\label{j5a}
\eeq
Thus if 
${\partial {\bf Q}\over \partial t}=\curl(\bfV\times \bfQ)$ and
$\div \bfQ=0$, then 
${D\over Dt}\int \bfQ\cdot d{\bf S} =0$.
Replacing $\bfQ$ with $\bfB$ we have the proof of flux freezing by direct differentiation.

The derivation above is just the surface integral analogue  of that used to prove Reynolds transport theorem. The latter describes the evolution of  a scalar volume integral over  a material volume that evolves in time from a velocity flow \cite{r03,mt,kc12}.

\section{Comparison  to other derivations}

Like the derivation above, the approaches in \cite{gp04} and \cite{roberts07}   do proceed by direct differentiation of the flux integral and  separately compute ${D{\bf B}/dt}$ and $D {d{\bf S}/dt}$ by considering the infinitesimal evolution of these quantities. However these approaches make no explicit mention of the Jacobian so  the connection to the basic method of surface integral transformations is not made explicit.  They  also assume $\div \bfB=0$ and do not  provide a physical interpretation of  why $\div\bfB\ne 0$ can violate flux freezing.

The derivation of the Kelvin circulation theorem in Ref. \cite{landau87}  also proceeds by direct
differentiation, but uses Stokes' theorem first so that the derivative is taken on the  line integral of velocity. 
The fact that vorticity is the curl of the velocity automatically ensures that vorticity is divergence-free.  Also,  if the same approach were applied to magnetic flux  then the mathematical analogue to the velocity in the proof is the vector potential which itself is not a gauge invariant quantity.  The derivation of section 1  avoids use of vector potential for the case that $\bfQ=\bfB$ and carries the divergence term to the very end.

Most noteworthy  is that   in addition to starting with $\div\bfB=0$,  common MHD presentations
 \cite{alfven43, chandra61,parker79, kulsrud05,bs05}
   do not cleanly  show how to calculate  the material derivative of the flux integral. 
Instead these  approaches  are characterized by the following:
 A   bounded open surface $C$ in the plasma   is considered to evolve in a small time $\delta t$ to a new  surface $C'$ by the action of differentiable velocity flows. 
Because $\div \bfB=0$, Gauss' theorem tells us that the total integrated flux at time $t+\delta t$  
 through the closed surface formed by $C$, $C'$   and the 
quasi-cylindrical  "side" connecting  $C$ and $C'$ is zero. 
Mathematically, this means
\beq
\int \div \bfB \ dV =0=
-\int_{C}{\bf B}({\bf r},t+\delta t )\cdot d{\bf S}
+\int_{C'}{\bf B}({\bf r},t +\delta t)\cdot d{\bf S}+
\int_{side}{\bf B}({\bf r},t +\delta t)\cdot d{\bf S}.
\label{3}
\eeq
The last term is
 \beq
\int_{side}{\bf B}({\bf r},t +\delta t)\cdot d{\bf S} =
\int_{C}{\bf B}({\bf r},t +\delta t)\cdot( d{\bf l}\times {\bf V}\delta t)
 = \int_{C}\delta t({\bf V}\times {\bf B}({\bf r},t +\delta t))\cdot d{\bf l},
 \label{4}
  \eeq
  where $d\bf l$ is the line element around the boundary of surface $C$.
The  differential  change  in  magnetic flux through $C$ as it evoles from $C$ to $C'$ is  
\beq
\delta \Phi_B=  \int_{C'} B({\bf r},t+\delta t)\cdot d{\bf S}-
\int_C B({\bf r},t)\cdot d{\bf S}.
\label{5}
\eeq
 Using (\ref{4}) and  (\ref{5}) to replace the last and penultimate terms of 
 (\ref{3}) respectively,  gives
\beq
{\delta \Phi\over \delta t}=\int_C {\partial {\bf B}\over \partial t} d{\bf S}
+\int_{C}({\bf V}\times {\bf B}({\bf r},t ))\cdot d{\bf l}
=\int_C {\partial {\bf B}\over \partial t} d{\bf S}
+\int_{C}\curl ({\bf V}\times {\bf B}({\bf r},t ))\cdot d{\bf S},
\label{15bb}
\eeq
where the last equality follows from use of Green's theorem and   the limit of small $\delta t$  has allowed replacement of   $  {B({\bf r},t+\delta t)-B({\bf r},t)\over \delta t}$ by ${\partial {\bf B}\over \partial t}  $.  Finally, by writing   $\delta \Phi /\delta t=D\Phi/Dt$ and using 
 Eq. (\ref{1}) for $\nu_M=0$,   Eq. (\ref{2})  obtains.

By comparison to the  method of the previous section, derivations 
along the lines of those which  follow Eq. (\ref{3})-(\ref{15bb}) do not as lucidly 
 separate of the material time derivative of the integrand from that  of the measure.   In addition, since the starting point assumes  $\div \bfB=0$, the reader is not provided with the opportunity to understand why  $\div \bfB \ne 0$ can violate flux freezing.

\section{Seeing the role of $\div \bfB$}

 The need for clarity on why  $\div \bfQ\ne 0$  can violate flux freezing is further evidenced by  the ambiguity of the derivation of the  Kelvin vorticity theorem of   Ref. \cite{choudhuri}.  
  There (in contrast to the approach of \cite{landau87}) it is implied   
that any vector $\bfQ$ satisfying 
${\partial {\bf Q}\over \partial t}-\curl(\bfV\times \bfQ)  =0$ obeys flux freezing
and that  the property that $\bf Q$ is the curl of some other 
function (i.e. that  $\div \bfQ=0$) is unnecessary to prove flux conservation.  
But this is incorrect, and seemingly results in Ref. \cite{choudhuri} from an ambiguity in distinguishing the partial and material derivative (compare Eq. 44 of Ref. \cite{choudhuri} to   Eqn. \ref{j3} above).
As seen in   Eq. (\ref{j5a})  above,  a finite divergence term would violate flux freezing even if the former condition is satisfied.   Note that the last term   in Eq. (\ref{j5a})  containing the divergence actually cancels
the hidden divergence term  within the penultimate term  since  $-\curl (\bfV \times \bfQ)$ includes
a term $-\bfV\nabla \cdot\bfQ$ when expanded with vector identities.  However, the ideal MHD magnetic induction
equation does not involve $\div \bfB=0$ in its derivation. Thus  for  $\bfQ=\bfB$, the first two terms on the right of (\ref{j5a}) cancel,  leaving  the divergence term whose physical meaning I now discuss.  

The divergence term of  Eq. (\ref{j5a})  represents the net advection of field line divergence through the evolving surface.  For $\bfQ=\bfB$, a finite $\div \bfB=4\pi \rho_m$ would imply the existence of magnetic monopoles of magnetic charge density $\rho_m$ by analogy to Gauss' law  for
 electric charge.  The magnetic field lines emanating from a  magnetic monopole have  a net  magnetic flux through any  spherical surface surrounding the monopole.  To see that a net advection  of  magnetic monopoles through a surface  would   change the magnetic flux through that surface,  first  consider the contribution form a single monopole of positive magnetic charge which has all field lines  directed radially outward. As the monopole approaches the surface from one side and passes through to the other, the sign of its contribution to the flux through that surface changes.
   An advection of a net density of monopoles of one sign through the surface would then by extension also
   change the flux through the surface with time.
   
Note that a $\div \bfB$  term in the flux evolution equation need   not  be the only consequence to MHD in a hypothetical plasma of  arbitrarily large   magnetic monopole  densities.  
 In the same way that we derive  the standard Ohm's law for MHD by subtracting electron and ion momentum density equations, we would  also have to derive a magnetic Ohm's law by subtracting positive and negative magnetic monopole charge density equations.   The two Ohm's laws would be coupled.
 Further study of magnetic monopoles is beyond the scope of the current paper.

\section{Conclusion}
Commonly used derivations of magnetic  flux freezing  tiptoe around showing how to take the material derivative of a   general flux integral. In addition, derivations which begin with $\div \bfB=0$ from the outset
do not provide the opportunity for a  physical understanding of why  $\div \bfB\ne 0$ could violate  flux freezing.  The direct  differentiation method of section 1 overcomes both of these shortcomings.

\ack
 I thank Jack Thomas and Kandu Subramanian for discussions, and Hal Cambier and Henk Spruit
 for comments and  acknowledge support from NSF grants AST-0406799
AST-0807363, and NSF Grant PHY-0903797.


\end{document}